\documentclass[twocolumn,pre,superscriptaddress,amsmath,amssymb,longbibliography]{revtex4}
\usepackage{graphicx}
\usepackage{float}
\usepackage{dcolumn}
\usepackage{xcolor}
\usepackage{latexsym,bm}
\usepackage{amsmath}
\usepackage{amsfonts}

\begin{document}



\title{Spectral Leakage and Masking Effects in the Measurement of Hyperuniformity}


\author{Yang Jiao}
\affiliation {Materials Science and Engineering, Arizona State University, Tempe, AZ 85287} \affiliation{Department of Physics, Arizona State University, Tempe, AZ 85287}

\date{\today}

\begin{abstract}
Disordered hyperuniformity is a recently discovered novel state of matter, characterized by a complete suppression of normalized infinite-wavelength density fluctuations as in perfect crystals and lack of conventional long-range order nor broken symmetry as in glasses. The detection of hyperuniformity relies critically on accurate characterization of the small-wavenumber behavior of the static structure factor of the system. In practice, however, measurements are performed on finite subsystems or through incomplete observations that effectively mask portions of the underlying configuration. Inspired by a recent numerical study [Y. Liu, X. Li, J. Tian, X. Yan, G. Zhang, {\it J. Chem. Phys.} {\bf 164}, 094102 (2026)], we develop a unified theoretical framework that quantifies how finite windows and spatially correlated binary masks modify the observed structure factor. We show that the measured structure factor $S_{obs}(k)$ is the convolution of the intrinsic structure factor with the spectral density of the observation function, whether it is a compact window or an extended random mask. For generic hyperuniform systems with small-$k$ scaling $S(k)\sim k^{\alpha}$, finite observation window induces a universal quadratic leakage term at sufficiently small wavenumbers (i.e., $k \lesssim 1/L$), leading to an apparent $k^{2}$ scaling independent of the true exponent. The true hyperuniform exponent $\alpha$ can only be measured in the intermediate regime $1/L \ll k \ll q_c$. In stealthy hyperuniform systems, where the intrinsic structure factor possesses a spectral gap, all observed small-$k$ power arises entirely from this convolution mechanism. For spatially correlated masks, we derive the corresponding convolution relation in terms of the mask spectral density and identify conditions under which hyperuniform signatures are suppressed, preserved, or distorted. Our results establish quantitative criteria for reliably extracting intrinsic scaling exponents and distinguishing genuine hyperuniform order from measurement-induced artifacts.
\end{abstract}


\maketitle

\section{Introduction}




Disordered hyperuniform (DHU) many-body systems possess a ``hidden order'' manifested as complete suppression of normalized infinite-wavelength density fluctuations like crystals, yet they lack conventional long-range order as in amorphous materials \cite{To03, To18a, torquato2026hyperuniformity}. This unique structural characteristics endows hyperuniform systems with unusual physical properties compared to their crystalline counterpart, such as high-degree of isotropy and robustness against defects \cite{Fl09, klatt2022wave, Zh16, Ch18a, torquato2021diffusion, Xu17, Le16, yu2023evolving, lee2026non, youn2026phase, maher2026probing, xu2026extrinsic, liang2025disordered, jiao2026acoustic}. Hyperuniformity is characterized by a local number variance $\sigma_N^2(R)$ associated with a spherical window of radius $R$ in $\mathbb{R}^d$ that grows more slowly than the window volume in the large-$R$ limit \cite{To03, To18a}, i.e., 
\begin{equation}
\lim_{r\rightarrow \infty} \sigma_N^2(R)/R^d = 0. 
\end{equation}
Equivalently, the static structure factor $S({\bf k})$ vanishes in the infinite-wavelength (or zero-wavenumber) limit, i.e., 
\begin{equation}
    \lim_{|{\bf k}|\rightarrow 0}S({\bf k}) = 0
\end{equation}
where ${\bf k}$ is the wavenumber. For statistically isotropic systems, the structure factor only depends on the wavenumber $k = |{\bf k}|$. The small-$k$ scaling behavior of $S(k)$, i.e., $S({k}) \sim k^\alpha$ determines the large-$R$ asymptotic behavior of $\sigma_N^2(R)$, based on which all DHU systems can be categorized into three classes \cite{To18a} (see Sec. II for details).


Recently, a wide spectrum of equilibrium \cite{To15, Ba09} and non-equilibrium \cite{Ga02, Do05, Za11a, maire2026hyperuniformity, maire2025hyperuniformity, wang2025hyperuniform, maire2025hyperuniform} many-body systems, in both classical \cite{Ku11, Hu12, Dr15, chen2021multihyperuniform, zhang2023approach, Ch18b, shang2026torque, bolton2026ideal, maher2026probing} and quantum mechanical \cite{Fe56, Ge19quantum, sakai2022quantum, Ru19, Sa19, Zh20, Ch21, nanotube, chen2025anomalous, vanoni2025effective, asakura2026impact, jeon2026quantum} varieties, have been identified to possess the property of disordered hyperuniformity. Specific examples include certain biological systems \cite{Ji14, Ma15, ge2023hidden, liu2024universal, li2025fluidization,tang2024tunable, hu2025causes}, driven non-equilibrium systems \cite{He15, Ja15, We15, salvalaglio2020hyperuniform, nizam2021dynamic, zheng2023universal, wang2025hyperuniformprl, leoni2025confinement, ballestero2025emergence}, active-particle fluids \cite{Le19, lei2019hydrodynamics, huang2021circular, zhang2022hyperuniform, oppenheimer2022hyperuniformity, backofen2024nonequilibrium, lei2023does, maire2025hyperuniform, maire2026hyperuniformity}, dynamic random organizing systems \cite{hexner2017noise, hexner2017enhanced, weijs2017mixing, wilken2022random}, and quantum material systems \cite{Ge19quantum, sakai2022quantum, chen2025anomalous,De16, chen2023disordered, Ch21,Zh21,sanchez2023disordered, oppenheimer2022hyperuniformity}. Notably, hyperuniformity has been discovered in a variety of experimental systems, including amorphous 2D materials \cite{Zh20, Ch21, PhysRevB.103.224102}, vortex matters \cite{sanchez2023disordered, oppenheimer2022hyperuniformity}, avian photoreceptors \cite{Ji14}, leaf vein networks \cite{liu2024universal}, and vegetation patterns in arid areas \cite{ge2023hidden, hu2025causes}, to name but a few.


\begin{figure}[ht]
\begin{center}
$\begin{array}{c}\\
\includegraphics[width=0.4\textwidth]{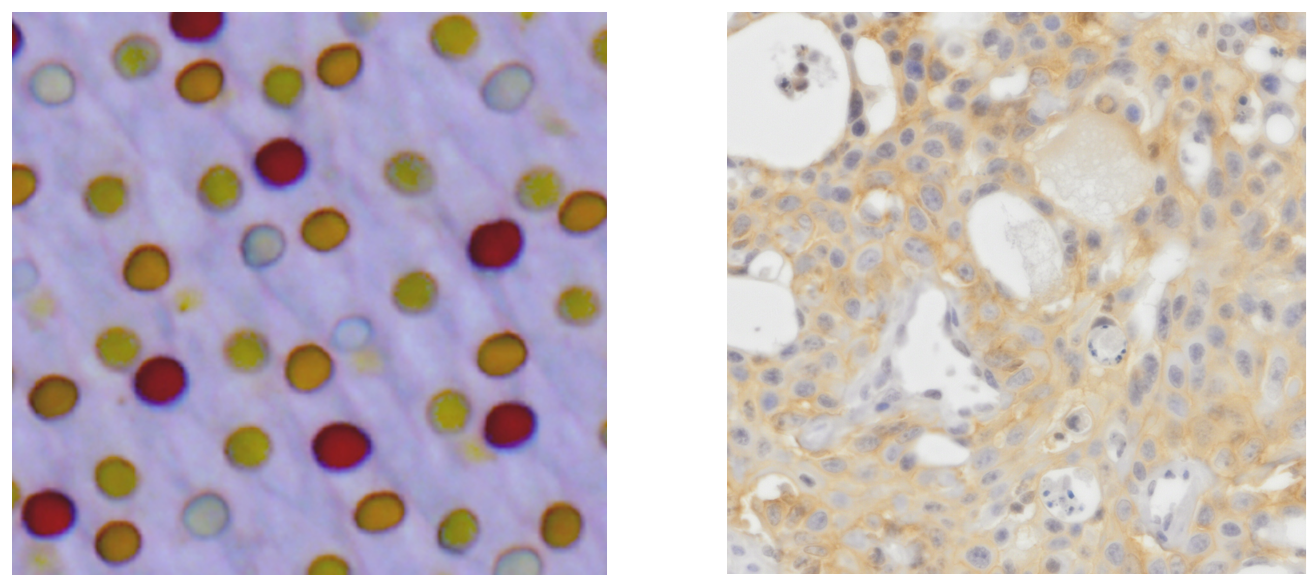}
\end{array}$
\end{center}
\caption{Examples of experimental systems in which hyperuniformity detection is based on measurements performed on finite subsystems (left panel) or effectively masked portions (right panel). Left panel: optical image of avian photoreceptor patterns where different types of receptors are shown in different colors. The linear size of the system is approximately 25 $\mu m$. Right panel: optical image of breast cancer cells (dark blue) in the stroma (light brown), where the stroma effectively masks the cancer cells. The linear szie of the system is approximately 150 $\mu m$.} \label{fig1}
\end{figure}

In the study of disordered hyperuniform systems, a key step is to ascertain the degree of hyperuniformity, which is typically based on measurements performed on finite subsystems or through incomplete observations that effectively mask portions of the underlying configuration, see Fig. \ref{fig1} for illustration. The detection of hyperuniformity relies critically on accurate characterization of the small-wavenumber behavior of the static structure factor of the system, which is essentially limited by the finite system size, especially in experimental systems. In a recent numerical study \cite{liu2025impact}, Liu et al. systematically examined how finite-window sampling and reciprocal-space radial binning affect the detection of hyperuniformity in disordered systems using both simulations and experimental data. They demonstrated that finite-window sampling typically preserves hyperuniform classification despite affecting the measured hyperuniformity exponents. In a previous study \cite{new2017}, Ikeda et al. highlighted the subtle challenges involved in interpreting large-scale density fluctuations and the importance of carefully distinguishing genuine long-range correlations from artifacts associated with finite windows and local structural effects. 



In this work, we develop a unified theoretical framework that describes how both finite windows and spatially correlated binary masks modify the observed structure factor. Within this framework, the measured structure factor $S_{\mathrm{obs}}(k)$ is expressed as the convolution of the intrinsic structure factor with the spectral density of the observation function, encompassing both compact windows and extended random masks. This convolution has important consequences for the interpretation of small-wavenumber behavior. In particular, for generic hyperuniform systems with $S(k)\sim k^{\alpha}$, finite observation induces a universal quadratic contribution at the lowest accessible wavenumbers (i.e., $k \lesssim 1/L$), which can give rise to an apparent $k^2$ scaling independent of the true exponent. As a result, the intrinsic scaling exponent $\alpha$ can only be reliably extracted within an intermediate regime $1/L \ll k \ll q_c$. In the case of stealthy hyperuniform systems, where the intrinsic structure factor exhibits a spectral gap, all observed small-$k$ fluctuations originate from this convolution mechanism. More generally, for spatially correlated masks, the interplay between the intrinsic structure factor and the mask spectral density determines whether hyperuniform signatures are preserved, suppressed, or distorted. These results provide a systematic basis for interpreting measurements and establish practical criteria for distinguishing intrinsic hyperuniform behavior from observation-induced artifacts.

The rest of the paper is organized as follows: In Sec.II, we provide definitions of preliminaries of hyperuniformity. In Sec. III, we provide the mathematical framework for finite-window effects on the measured structure factor. In Sec. IV, we derive the framework for the effects of random binary masks on the measured structure factor and provide concrete numerical examples. In Sec. V, we provide concluding remarks and discuss implications of our work.



\section{Definitions and Preliminaries}



Consider a statistically homogeneous point configuration in $d$-dimensional Euclidean space $\mathbb{R}^d$ with
microscopic density
\begin{equation}
\rho(\mathbf{r}) = \sum_{j=1}^N \delta(\mathbf{r}-\mathbf{r}_j).
\end{equation}
A point configuration is completely characterized by an infinite set of $n$-point correlation functions $\rho_n(\mathbf{r}_1,\dots,\mathbf{r}_n)$, each of which is proportional to the probability of finding $n$ points at the positions $\mathbf{r}_1,\dots,\mathbf{r}_n$ \cite{To02a}.
For statistically homogeneous systems, $\rho_1(\mathbf{r}_1)=\rho$, and $\rho_2(\mathbf{r}_1,\mathbf{r}_2)=\rho^2 g_2(\mathbf{r})$, where $\mathbf{r}=\mathbf{r}_1-\mathbf{r}_2$, and $g_2(\mathbf{r})$ is the pair correlation function.
If the system is also statistically isotropic, then $g_2(\mathbf{r})=g_2(r)$, where $r=|{\bf r}|$.
The ensemble-averaged structure factor $S(\mathbf{k})$ is defined as
\begin{equation}
    S(\mathbf{k})=1+\rho\Tilde{h}(\mathbf{k})
\end{equation}
where $\Tilde{h}(\mathbf{k})$ is the Fourier transform of the total correlation function $h(\mathbf{r})=g_2(\mathbf{r})-1$, and ${\bf k}$ is the wave vector.

For a single periodic point configuration with $N$ particles at positions $\mathbf{r}^N = (\mathbf{r}_1,\dots,\mathbf{r}_N)$ within a fundamental cell $F$ of a lattice $\Lambda$, the scattering intensity $\mathbb{S}(\mathbf{k})$ is given by
\begin{equation}\label{eq:Skcomp}
    \mathbb{S}(\mathbf{k}) = \frac{|\sum_{j=1}^N \textrm{exp}(-i\mathbf{k}\cdot\mathbf{r}_j)|^2}{N}.
\end{equation}
In the thermodynamic limit, the scattering intensity of an ensemble of an $N$-particle configurations in $F$ is related to $S(\mathbf{k})$ by
\begin{equation}
    \lim_{N,V_F\rightarrow\infty}\langle \mathbb{S}(\mathbf{k})\rangle = (2\pi)^d \rho \delta(\mathbf{k}) + S(\mathbf{k}),
\end{equation}
where $\rho = N/V_F$, $V_F$ is the volume of the fundamental cell, and $\delta$ is the Dirac delta function \cite{To03}.
For finite-$N$ simulations under periodic boundary conditions, Eq.~(\ref{eq:Skcomp}) is used to compute $S(\mathbf{k})$ directly by averaging over configurations. The smallest vector number that can be accessed of a finite system is $k = |{\bf k}| = 2\pi/L$, where $L$ is the linear size of the system.


Consider systems characterized by a structure factor with a radial power law in the vicinity of the origin,
\begin{equation}
    S(\mathbf{k})\sim|\mathbf{k}|^{\alpha}\;\textrm{for}\;|\mathbf{k}|\rightarrow0.
\end{equation}
The exponent $\alpha$ is referred to as the hyperuniformity exponent. For {\it hyperuniform} systems, $\alpha > 0$. A (standard) {\it nonhyperuniform} system has $\alpha = 0$, i.e., S({\bf k}) approaches a non-zero constant in the zero-wavenumber limit. An {\it antihyperuniform} system is one possessing a diverging S({\bf k}) in the zero-wavenumber limit, i.e., with $\alpha < 0$. For hyperuniform systems, $\alpha$ determines large-$R$ scaling behaviors of the number variance \cite{To03, To18a}, according to which all hyperuniform systems can be categorized into three different classes:
\begin{equation}\label{eq:classes}
    \sigma^2_N(R)\sim
    \begin{cases}
    R^{d-1}&\alpha > 1, \textrm{class I}\\
    R^{d-1}\textrm{ln}(R)&\alpha = 1, \textrm{class II}\\
    R^{d-\alpha}&\alpha < 1, \textrm{class III}.
    \end{cases}
\end{equation}
Classes I and III are the strongest and weakest forms of hyperuniformity, respectively. 


Stealthy hyperuniform systems are a special subset of class-I hyperuniform systems possessing a zero structure factor for a range of wavevectors around the origin, i.e.,
\begin{equation}
S({\bf k}) = 0, \quad\text{for}\quad {\bf k} \in \Omega,
\label{eq_stealthy}
\end{equation}
excluding the forward scattering. Stealthy hyperuniform systems include all crystals, most quasicrystals and certain special disordered systems \cite{To18a}. The degree of stealthiness and short-range order in the system is determined by a tuning parameter $\chi$ measuring the fraction of the independently constrained degrees of freedom $M$ within the exclusion region $\Omega$ (i.e., half the number of ${\bf k}$ points in $\Omega$) \cite{To15, Zh15a, Zh15b, Zh17}, i.e.,
\begin{equation}
    \chi = \frac{M}{(N-1)d}
    \label{eq_chi_ratio}
\end{equation}
where $N$ is the total number of points in the systems. We note that $d$ degrees of freedom associated with the trivial overall translation of the entire system are subtracted in Eq.~(\ref{eq_chi_ratio}). Stealthy hyperuniform systems distinguish themselves from standard hyperuniform systems in that they completely suppress density fluctuations at the intermediate to infinite wavelengths. In contrast, standard hyperuniform systems only completely suppress infinite-wavelength density fluctuations, i.e., in the zero-$|{\bf k}|$ limit. \\


\section{Finite-Window Effects on the Measured Structure Factor}

\subsection{Exact Convolution Formula}

\begin{figure}[ht]
\begin{center}
$\begin{array}{c}\\
\includegraphics[width=0.3\textwidth]{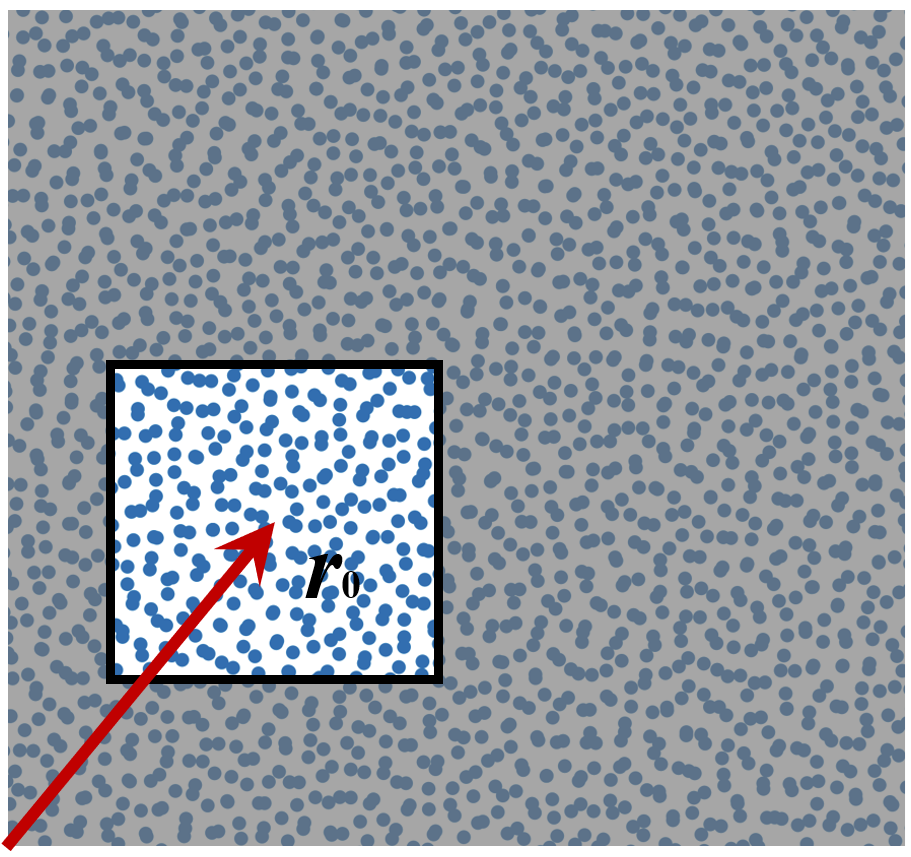}
\end{array}$
\end{center}
\caption{Illustration of a finite observation window revealing only a subset of the system based on which hyperuniformity detection is performed.} \label{fig2}
\end{figure}

Suppose we measure the structure factor from a finite binary observation window of linear size $L$ centered at $\mathbf r_0$ (see Fig. \ref{fig2}) defined as
\begin{equation}
W_{\mathbf r_0}(\mathbf r)
=
W(\mathbf r-\mathbf r_0),
\end{equation}
where $W(\mathbf r)$ is a reference window centered at the origin. For example, in two dimensions, a square window possesses
\begin{equation}
W(\mathbf r)=
\begin{cases}
1, & |x|<L/2,\ |y|<L/2, \\
0, & \text{otherwise},
\end{cases}
\end{equation}
and a circular window possesses
\begin{equation}
W(\mathbf r)=
\begin{cases}
1, & |\mathbf r|<L, \\
0, & \text{otherwise}.
\end{cases}
\end{equation}

The observed (windowed) density is then given by:
\begin{equation}
\rho_{\mathrm{sub}}(\mathbf r)
=
\rho(\mathbf r)\,W_{\mathbf r_0}(\mathbf r).
\end{equation}
Multiplication in real space implies convolution in Fourier space:
\begin{equation}
\tilde\rho_{\mathrm{sub}}(\mathbf k)
=
\int\frac{d^d q}{(2\pi)^d}\,
\tilde\rho(\mathbf q)\,
\widetilde{W_{\mathbf r_0}}(\mathbf k-\mathbf q).
\end{equation}
Substituting the translation identity
\begin{equation}
\widetilde{W_{\mathbf r_0}}(\mathbf k)
=
e^{-i\mathbf k\cdot\mathbf r_0}\,
\widetilde W(\mathbf k),
\end{equation}
we obtain
\begin{equation}
\tilde\rho_{\mathrm{sub}}(\mathbf k)
=
\int\frac{d^d q}{(2\pi)^d}\,
\tilde\rho(\mathbf q)\,
e^{-i(\mathbf k-\mathbf q)\cdot\mathbf r_0}
\widetilde W(\mathbf k-\mathbf q).
\end{equation}
where $\tilde{W}$ is the Fourier transform of the window. Averaging over random window placements ${\bf r}_0$ removes phase correlations and yields the exact relation for the observed structure factor:
\begin{equation}
S_{\mathrm{obs}}(\mathbf{k})
=
\int \frac{d^d q}{(2\pi)^d}
\, S(\mathbf{q})
\, |\tilde{W}(\mathbf{k}-\mathbf{q})|^2.
\label{eq:convolution}
\end{equation}
Thus, the measured structure factor $S_{\mathrm{obs}}(\mathbf{k})$ is the convolution of the true structure factor $S({\bf k})$ with the window power spectrum.

For a window of linear size $L$, we re-write the Fourier transform of the window function as 
\begin{equation}
\tilde{W}(\mathbf{k}) = L^{d} w(L\mathbf{k}),
\end{equation}
where $w$ is an $L$-independent shape function (e.g., a product of sinc
functions for a square window). Substituting the above into Eq.~(\ref{eq:convolution}) gives
\begin{equation}
S_{\mathrm{obs}}(\mathbf{k})
=
\frac{L^{d}}{(2\pi)^d}
\int d^d u \,
S\!\left(\frac{\mathbf{u}}{L}\right)
|w(L\mathbf{k}-\mathbf{u})|^2.
\label{eq:scaled}
\end{equation}
where the dimensionless wavevector ${\bf u}= L{\bf q}$, which makes the $L$-scaling explicit in Eq. {\ref{eq:scaled}}.

\subsection{Generic Hyperuniform Case: $S(k) \sim k^\alpha$}

Assume a statistically isotropic parent system is hyperuniform with
\begin{equation}
S(k) \sim C k^{\alpha},
\qquad \alpha > 0,
\end{equation}
for $k$ below a crossover scale $k_c$. For sufficiently large $L$, one may substitute the small-$k$ form of $S(k)$ into
Eq.~(\ref{eq:scaled}), yielding
\begin{equation}
S_{\mathrm{obs}}(k)
\approx
\frac{C L^{d-\alpha}}{(2\pi)^d}
\int d^d u \,
|\mathbf{u}|^{\alpha}
|w(L\mathbf{k}-\mathbf{u})|^2.
\label{eq_explicit}
\end{equation}

We note for such systems, two asymptotic regimes emerge. We first discuss the {\it spectral leakage regime}, i.e., $k \ll 1/L$. Expanding Eq. (\ref{eq_explicit}) for small $Lk$,
\begin{equation}
S_{\mathrm{obs}}(k)
=
A_0 L^{d-\alpha}
+
A_2 L^{d-\alpha+2} k^2
+
O(k^4),
\end{equation}
where $A_0$ and $A_2$ are positive constants that depend on the window shape, i.e., 
\begin{equation}
    A_0 \;=\; \frac{C}{(2\pi)^d}\int_{\mathbb R^d} d^d u\; |\mathbf u|^{\alpha}\;|w(\mathbf u)|^2
\end{equation}
and 
\begin{equation}
    A_2 \;=\; \frac{C}{(2\pi)^d}\;\frac{1}{2d}\;
\int_{\mathbb R^d} d^d u\; |\mathbf u|^{\alpha}\;\Delta\!\bigl(|w(\mathbf u)|^2\bigr)
\end{equation}
We note that when computing the structure factor from a system with linear size $L$, the smallest discrete wavenumber is
\begin{equation}
    k_{min} \approx \frac{2\pi}{L}
\end{equation}
so strictly speaking one never probes $k \ll 1/L$ when computing the discrete structure factor of a subsystem of size $L$. However, the quadratic leakage regime does not require $k\ll1/L$. The condition $Lk\ll1$ in the asymptotic derivation is a sufficient condition for the Taylor expansion, but in practice the leakage behavior extends into the first few accessible modes with $k \sim O(1/L)$. In this context, the spectral leakage has two major effects: degrading the degree of hyperuniformity through a positive $S_{\mathrm{obs}}(k\rightarrow 0) = A_0 L^{d-\alpha}$ and a shift of scaling towards $k^2$.



We now investigate the regime $1/L \ll k \ll k_c$ so that $Lk\gg1$ but $k$ is still small enough that $S(k)\propto k^{\alpha}$ follows the asymptotic law at the relevant scales. For large argument $|L\mathbf k|\gg1$, the function $|w(L\mathbf k-\mathbf u)|^2$ in Eq. (\ref{eq_explicit}) is sharply localized in $\mathbf u$ around $\mathbf u\approx L\mathbf k$, i.e. it acts as an $O(1)$-width kernel in the $\mathbf u$ variable centered at $\mathbf u=L\mathbf k$. Change integration variable $\mathbf u = L\mathbf k + \mathbf v$ to obtain
\begin{equation}
S_{\mathrm{obs}}(\mathbf k)
=\frac{L^{d}}{(2\pi)^d}\int_{\mathbb R^d} d^d v\;
S\!\Big(\frac{L\mathbf k+\mathbf v}{L}\Big)\;|w(\mathbf v)|^2.
\label{eq:shifted}
\end{equation}
where we have used $|w(-\mathbf v)|^2=|w(\mathbf v)|^2$. Because $\mathbf v$ varies over $O(1)$ values while $L$ is large, expand the slowly varying function $S(\mathbf k+\mathbf v/L)$ in powers of $\mathbf v/L$, and insert the expansion into Eq.~\eqref{eq:shifted} and use the parity of $|w(\mathbf v)|^2$ (even in $\mathbf v$) to drop all odd terms yields: 
\begin{equation}
S_{\mathrm{obs}}(\mathbf k)
\approx
\frac{L^{d}}{(2\pi)^d}\;S(\mathbf k)\;\int_{\mathbb R^d} d^d v\;|w(\mathbf v)|^2,
\label{eq:leading-local}
\end{equation}
where we defined the window power integral
\begin{equation}
    W_2\equiv \int_{\mathbb R^d} |w(\mathbf v)|^2\,d^d v.
\end{equation}
Thus, to leading order in the localization approximation,
\begin{equation}
S_{\mathrm{obs}}(\mathbf k)
\simeq \frac{L^{d}W_2}{(2\pi)^d}\; S(\mathbf k)
\qquad\text{for }  \frac{1}{L}\ll k \ll q_c.
\label{eq:intrinsic-box}
\end{equation}
This result indicates when $Lk \gg 1$, the convolution kernel is localized and
\begin{equation}
S_{\mathrm{obs}}(k)
\approx
\tilde{C}\, k^{\alpha},
\label{eq_intrinsic}
\end{equation}
recovering the true hyperuniform scaling (up to a normalization factor). Similar to the spectral leakage regime, although the analysis assumes $Lk \gg 1$, in practice we expect to observe Eq. (\ref{eq_intrinsic}) for $Lk > 1$. Since $k_{min} \approx 2\pi/L$, as one moves away from the first few $k$ points, the scaling behavior of $S(k)$ should exhibit the intrinsic scaling $S(k) \sim k^{\alpha}$. We refer to this regime as the {\it intrinsic regime}.

In practical computations, one usually reports an intensive structure factor normalized by the observed particle number $N_{\mathrm{sub}}\propto \rho\,\phi\,L^{d}$ (where $\rho$ is the parent number density and $\phi$ the window/ mask filling fraction). Dividing both sides of Eq.~\eqref{eq:intrinsic-box} by $N_{\mathrm{sub}}$ removes the explicit $L^{d}$ factor and yields a $k$--independent multiplicative prefactor:
\begin{equation}
S_{\mathrm{obs}}^{\mathrm{(int)}}(\mathbf k)
\equiv \frac{1}{N_{\mathrm{sub}}}\big\langle|\tilde\rho_{\mathrm{obs}}(\mathbf k)|^2\big\rangle
\simeq \frac{W_2}{(2\pi)^d \rho\,\phi}\; S(\mathbf k).
\end{equation}
Hence the scaling behavior of $S(k)$ is preserved in the intrinsic regime; only the amplitude is rescaled by a constant determined by the window shape and normalization.

The first nonzero correction arises from the quadratic term in the expansion:
\begin{equation}
\Delta S_{\mathrm{obs}}(\mathbf k)
\approx
\frac{L^{d}}{(2\pi)^d}\,\frac{1}{2L^2}\,\partial_a\partial_b S(\mathbf k)\,
\int d^d v\; v_a v_b\,|w(\mathbf v)|^2.
\end{equation}
For an isotropic window one may write $\int v_a v_b|w|^2 = \delta_{ab} M_2/d$ with
\begin{equation}
M_2\equiv \int_{\mathbb R^d} |\mathbf v|^2 |w(\mathbf v)|^2\,d^d v,
\end{equation}
so the relative correction scales as
\begin{equation}
\frac{\Delta S_{\mathrm{obs}}(\mathbf k)}{S_{\mathrm{obs}}(\mathbf k)}
\sim \frac{1}{L^2}\;\frac{\partial^2 S(\mathbf k)}{S(\mathbf k)} \sim O\!\Big(\frac{1}{(L k)^2}\Big),
\end{equation}
using that for a power law $S(k)\sim k^{\alpha}$ the second derivative is $\partial^2 S\sim k^{\alpha-2}$. Therefore the localization approximation is accurate provided $L k \gg 1$, and the fractional error decays as $(Lk)^{-2}$. In practice one therefore requires $k$ a few times larger than $1/L$ for the intrinsic scaling to be observed with small relative error. Under these conditions the convolution reproduces the intrinsic power law $S(k)\propto k^{\alpha}$ up to a $k$-independent prefactor; the leading relative error is $O((Lk)^{-2})$.



\begin{figure}[ht]
\begin{center}
$\begin{array}{c}\\
\includegraphics[width=0.48\textwidth]{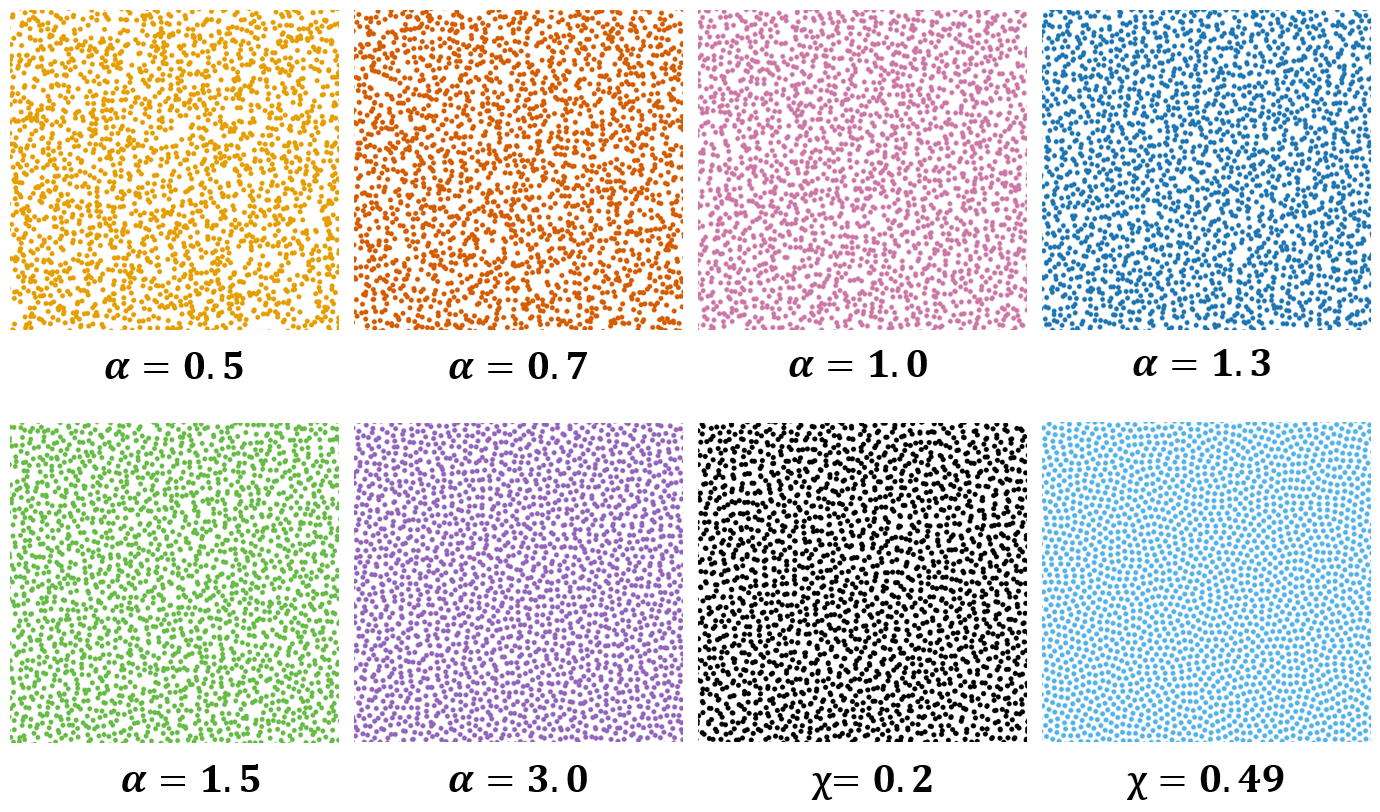}
\end{array}$
\end{center}
\caption{Representative disordered hyperuniform point configurations in $\mathbb{R}^2$ with varying hyperuniformity exponent $\alpha = 0.5$, 0.7, 1.0, 1.3, 1.5, 3.0, and stealthy hyperuniform systems with $\chi = 0.2$ and 0.49, spanning from class I to class III hyperuniform systems.} \label{fig3}
\end{figure}

To verify these results, we numerically computed the structure factor of a variety of disordered hyperuniform point configurations in $\mathbb{R}^2$ with $N = 10~000$ in a periodic unit square box with varying hyperuniformity exponent $\alpha = 0.5$, 0.7, 1.0, 1.3, 1.5, and 3.0, spanning from class I to class III hyperuniform systems \cite{liu2025impact}, see Fig. \ref{fig3}. For each system, we consider square windows with edge length $L = 0.1$, 0.5 and 1.0; for each size, 10 independent configurations are used. The scaling exponents are estimated using $k \in [2, 6]k_{min}$, representing the intrinsic regime. The results are shown in Table I. We observe that increasing observation window size $L$ generally improves the accuracy of numerically estimated $\alpha$ values, consistent with results reported in Ref. \cite{liu2025impact}.


\begin{table}[htb]
\centering
\caption{Estimated hyperuniform exponent $\alpha$ of a variety of disordered hyperuniform systems from finite observation windows of varying sizes.}
\label{tab:scalings}
\begin{tabular}{c|c|c|c|c}
\toprule
 & gro. tru. & $L = 0.1$ & $L = 0.5$ & $L = 1.0$ \\ \hline
Class III & $0.5$ & 0.47$\pm$0.04 & 0.48$\pm$0.04 & 0.51$\pm$0.02  \\\hline
Class III & $0.7$ & 0.69$\pm$0.06 & 0.68$\pm$0.04 & 0.71$\pm$0.03  \\\hline
Class II & $1.0$ & 0.96$\pm$0.05 & 0.96$\pm$0.06 & 0.99$\pm$0.03  \\\hline
Class I & $1.3$ & 1.24$\pm$0.08 & 1.28$\pm$0.09 & 1.31$\pm$0.06  \\\hline
Class I & $1.5$ & 1.33$\pm$0.12 & 1.45$\pm$0.07 & 1.48$\pm$0.06  \\ \hline
Class I & $3.0$ & 2.52$\pm$0.18 & 2.71$\pm$0.15 & 3.02$\pm$0.09  \\
\hline
Steal. $\chi = 0.2$ & $\infty$ & 1.93$\pm$0.07 & 1.97$\pm$0.07 &  - \\
\hline
Steal. $\chi = 0.49$ & $\infty$ & 1.96$\pm$0.08 & 2.02$\pm$0.05 &  - \\
\hline
\hline
\end{tabular}
\end{table}


For fixed observation parameters (window linear size $L$, average subsystem particle number $N_{\rm sub}$, and sampling scheme), the accuracy of fitted small-$k$ exponents depends strongly on the true exponent $\alpha$. For example, systems with small $\alpha$, the estimates are very close to the ground-truth value. While exponent estimation for strongly hyperuniform systems with larger exponents ($\alpha \gtrsim 2$) exhibit systematic downward bias. This is mainly because the intrinsic signal $S(k)\sim k^\alpha$ decays rapidly at small $k$, making it increasingly susceptible to contamination by leakage and additive noise. 

In particular, the measured structure factor may be decomposed as
\[
S_{\rm obs}(k)=C k^\alpha + S_{\rm leak}(k) + S_{\rm noise}(k),
\]
where $S_{\rm leak}(k)\simeq S_0+B k^2$ is the window-induced leakage and $S_{\rm noise}$ is the sampling/shot-noise floor. The intrinsic signal dominates only for
\[
k \gtrsim k_\times \equiv \bigg(\frac{S_0+S_{\rm noise}}{C}\bigg)^{1/\alpha}.
\]
Because $k_\times$ grows with $\alpha$ (for fixed $S_0,S_{\rm noise},C$), high-$\alpha$ systems require larger accessible $k$ (or smaller noise floors) before the intrinsic power law can be fitted reliably. If the fit window includes $k\lesssim k_\times$, the additive floor biases log–log fits toward lower slopes; this is the expected source of the underestimation observed for $\alpha=2$ and $\alpha=3$ in our numerics. Increasing observation window size generally improves the estimate.

\subsection{Stealthy Hyperuniform Case}

For stealthy hyperuniform systems, we have
\begin{equation}
S(k) = 0
\quad \text{for } |\mathbf{k}| < K.
\end{equation}
Inserting the above $S(k)$ into Eq.~(\ref{eq:convolution}),
we see that all observed small-$k$ power arises from convolution with
modes at $q \gtrsim K$:
\begin{equation}
S_{\mathrm{obs}}(k)
=
\int_{|\mathbf{q}| \gtrsim K}
\frac{d^d q}{(2\pi)^d}
\, S(\mathbf{q})
\, |\tilde{W}(\mathbf{k}-\mathbf{q})|^2.
\end{equation}
Expanding for $k \ll K$ again gives
\begin{equation}
S_{\mathrm{obs}}(k)
=
B_0
+
B_2 k^2
+
O(k^4).
\end{equation}
Thus, even though the parent spectrum has a hard gap, finite-window measurements generate a nonzero constant term (degrading the degree of hyperuniformity) and
a universal quadratic correction. We note these analyses are consistent with the theoretical arguments provided in Ref. \cite{liu2025impact}. Table I shows the numerically obtained estimates of $\alpha$ for stealthy hyperuniform systems with varying $\chi$, verifying our theoretical predictions.

\section{Effect of a binary mask on the measured structure factor}


\begin{figure}[ht]
\begin{center}
$\begin{array}{c}\\
\includegraphics[width=0.3\textwidth]{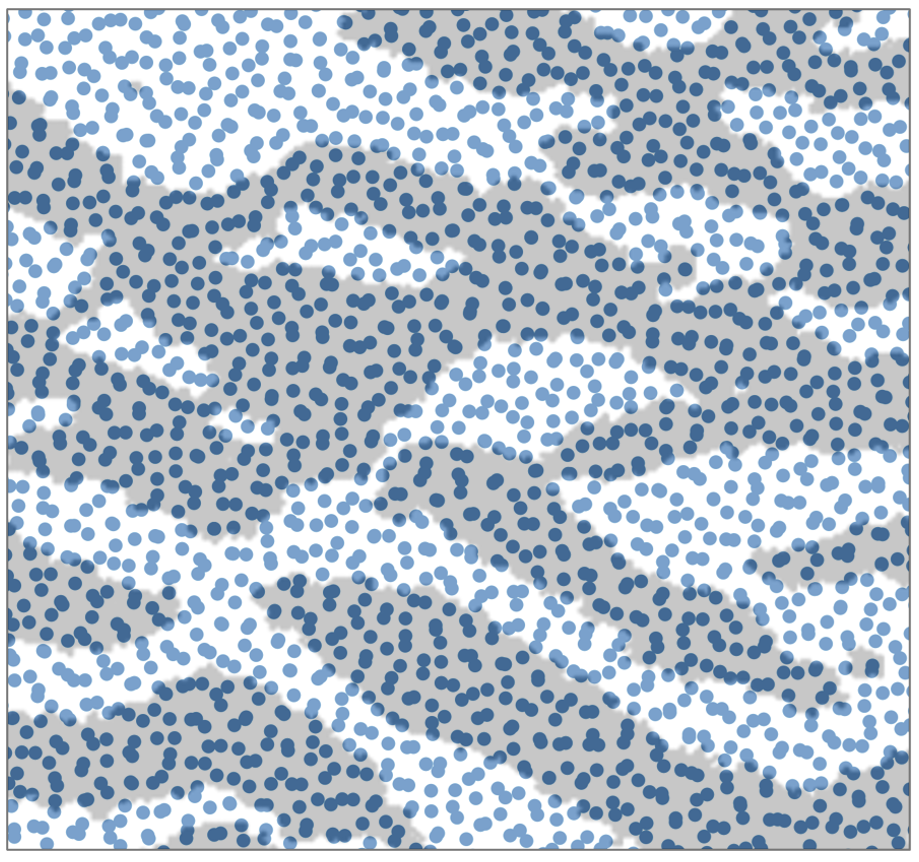}
\end{array}$
\end{center}
\caption{Illustration of a masked system where only a subset of points are revealed, based on which hyperuniformity detection is performed.} \label{fig4}
\end{figure}

Consider applying a binary mask \(M(\mathbf r)\in\{0,1\}\) across the entire system, i.e., 
\begin{equation}
M(\mathbf r)=
\begin{cases}
1, & \mathbf r \in \mathcal{V}, \\
0, & \text{otherwise}.
\end{cases}
\end{equation}
so that only points with \(M(\mathbf r)=1\) (falling into region $\mathcal{V}$) are observed (see Fig. \ref{fig4}). The observed (masked) density and its Fourier transform are respectively
\begin{equation}
\rho_{\mathrm{obs}}(\mathbf r) =M(\mathbf r)\,\rho(\mathbf r),
\end{equation}
and
\begin{equation}
\tilde\rho_{\mathrm{obs}}(\mathbf k)
\;=\;\int\frac{d^d q}{(2\pi)^d}\;\tilde\rho(\mathbf q)\,\tilde M(\mathbf k-\mathbf q),
\label{eq_rho_mask}
\end{equation}
i.e. multiplication in real space gives convolution in Fourier space. Here \(\tilde M(\mathbf k)\) is the Fourier transform of \(M(\mathbf r)\).


\subsection{Exact identity for the masked spectrum}

In order to obtain $S_{\mathrm{obs}}(k)$, we square the transform (\ref{eq_rho_mask}) and (optionally) average over independent ensembles of points and mask realizations. If the mask and the point process are {\it statistically independent}, the average factorizes and one obtains
\begin{equation}
\big\langle |\tilde\rho_{\mathrm{obs}}(\mathbf k)|^2\big\rangle
=
\int\frac{d^d q}{(2\pi)^d}\;
\big\langle |\tilde\rho(\mathbf q)|^2\big\rangle
\;\big\langle |\tilde M(\mathbf k-\mathbf q)|^2\big\rangle .
\label{eq:raw-conv}
\end{equation}
Introduce the mask \emph{spectral density} (per unit volume)
\begin{equation}
\chi_M(\mathbf p) \equiv \lim_{V\to\infty}\frac{\big\langle |\tilde M(\mathbf p)|^2\big\rangle}{V},
\end{equation}
and recall \(\big\langle|\tilde\rho(\mathbf q)|^2\big\rangle=(2\pi)^d N S(\mathbf q)\) (consistent with our normalization), Eq.~\eqref{eq:raw-conv} becomes
\begin{equation}
\big\langle |\tilde\rho_{\mathrm{obs}}(\mathbf k)|^2\big\rangle
= N V \int\frac{d^d q}{(2\pi)^d}\; S(\mathbf q)\;\chi_M(\mathbf k-\mathbf q).
\label{eq:conv-chi}
\end{equation}

The observed structure factor is given by 
\begin{equation}
   S_{\mathrm{obs}}(\mathbf k)\equiv \big\langle |\tilde\rho_{\mathrm{obs}}(\mathbf k)|^2\big\rangle/N_{\mathrm{obs}} 
\end{equation}
where $N_{\mathrm{obs}}=\phi N$
is the \emph{mean observed particle number}\, and
\(\phi=\langle M(\mathbf r)\rangle\) is the area/volume fraction of the mask (associated with region $\mathcal{V}$), 
we obtain the central relation
\begin{equation}
S_{\mathrm{obs}}(\mathbf k)
\;=\;
\frac{V}{\phi}\;
\int\frac{d^d q}{(2\pi)^d}\; S(\mathbf q)\; \chi_M(\mathbf k-\mathbf q).
\label{eq:boxed}
\end{equation}
Eq.~\eqref{eq:boxed} is the generalization of the finite-window convolution (where \(|\widetilde W|^2\) appeared) to the case of an \emph{extended correlated mask}. The correlation of the mask is encoded in the associated mask spectral density $\chi_M$, which is also related to the two-point correlation function of the mask. We thus show the observed structure factor with the mask is the convolution of the true structure factor with the mask spectral density, up to the dilution prefactor \(V/\phi\) coming from our normalization to \(N_{\mathrm{obs}}\).

\subsection{Case studies}

\subsubsection{Deterministic large mask (fixed pattern)}

If the mask \(M(\mathbf r)\) is a fixed deterministic pattern (not averaged over realizations), the mask spectral density \(\chi_M(\mathbf k)=|\tilde M(\mathbf k)|^2/V\) is a deterministic function. Eq.~\eqref{eq:boxed} reduces to the familiar convolution with \(|\tilde M|^2\):
\begin{equation}
    S_{\mathrm{obs}}(\mathbf k)=\frac{1}{\phi}\int\frac{d^d q}{(2\pi)^d}S(\mathbf q)\,|\tilde M(\mathbf k-\mathbf q)|^2 .
\end{equation}
This is the direct analogue of the finite-window result, but with the mask power spectrum replacing \(|\tilde W|^2\).


\subsubsection{Independent Bernoulli thinning (uncorrelated mask)}

\begin{figure}[ht]
\begin{center}
$\begin{array}{c}\\
\includegraphics[width=0.45\textwidth]{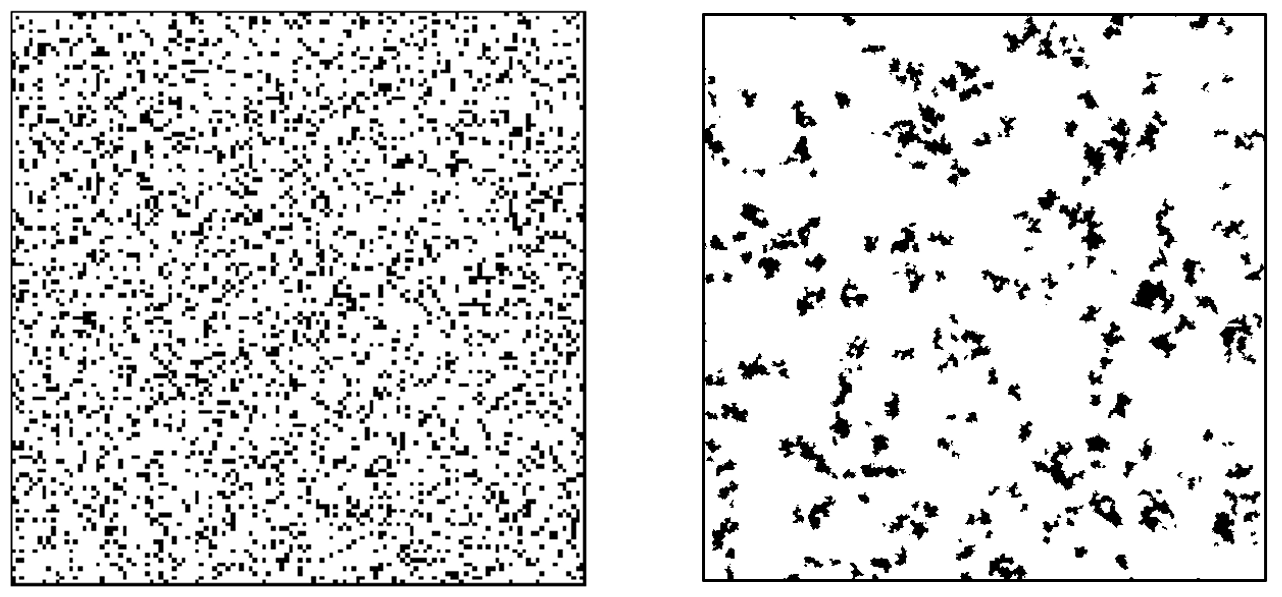}
\end{array}$
\end{center}
\caption{Illustration of a Bernoulli mask (left) and a Debye mask (right) with volume fraction $\phi = 0.15$.} \label{fig5}
\end{figure}

We now consider spatially uncorrelated masks, i.e., each spatial location in $\mathcal{V}$ possessing value 1 is  independent of one another with probability \(\phi\) (classical random thinning), see Fig. \ref{fig5} left panel. Its spectral density is
\begin{equation}
\chi_M(\mathbf k)=\phi(1-\phi) + (2\pi)^d\phi^2\delta(\mathbf k),
\end{equation}
where the \(\delta\)-term encodes the mean \(\phi\) (the constant background) and the constant term \(\phi(1-\phi)\) encodes white (uncorrelated) fluctuations of the mask. Inserting this into Eq.~\eqref{eq:boxed} and using \(\int (d^dq/(2\pi)^d)S(\mathbf q)= 1\) (consistency of normalization) yields
\begin{equation}
S_{\mathrm{obs}}(\mathbf k) = \phi S(\mathbf k)
\;+\; \frac{V}{\phi}\,\phi(1-\phi),
\end{equation}
where the first term is the attenuated parent spectrum, and the second term is the additive white noise.
After simplifying the prefactors (and passing to intensive normalization) this is commonly written as an attenuation of the parent spectrum plus an additive \(k\)-independent noise term: independent thinning both reduces the signal amplitude and injects white noise that lifts the small-\(k\) value.

\subsubsection{Mask with large-scale correlations}

If the mask has substantial power at small \(\mathbf k\) (i.e. \(\chi_M(\mathbf k)\) large near \(\mathbf k=\mathbf0\)), then Eq.~\eqref{eq:boxed} shows that small-\(|\mathbf k|\) behavior of \(S_{\mathrm{obs}}(\mathbf k)\) will be dominated by a convolution of the mask spectral weight near \(\mathbf 0\) with the parent \(S(\mathbf k)\) at similar scales. In particular, a mask with strong low-\(|\mathbf k|\) power will \emph{create} or \emph{enhance} apparent low-\(|\mathbf k|\) power in \(S_{\mathrm{obs}}\), potentially destroying an observed signature of hyperuniformity. Conversely, a mask whose \(\chi_M({\mathbf k})\) vanishes (or is small) near \(\mathbf k=0\) suppresses leakage into small \(|\mathbf k|\).

Here we consider a Debye mask (see Fig. \ref{fig5} right panel), which is characterized by an exponentially decaying correlation function, i.e., 
\begin{equation}
    S_2(r) = \phi(1-\phi)\exp(-r/a) + \phi^2
\end{equation}
where $a$ is a correlation length. The associated spectral density function is given by 
\begin{equation}
\chi_{M}({k}) = \phi_1 \phi_2 \frac{4a^2}{[1+(ka)^2]^{3/2}}  + (2\pi)^d\phi^2\delta(\mathbf k),
\label{eq_chi_Debye}
\end{equation}
where the \(\delta\)-term encodes the mean \(\phi\) (the constant background) and the first term encodes non-trivial correlations induced by the mask. In the small-$k$ limit, one has
\begin{equation}
\lim_{ k\rightarrow 0} \tilde \chi_{_V}({k}) = 4\phi_1\phi_2 a^2 + (2\pi)^d\phi^2\delta(\mathbf 0)
\end{equation}
which contains a non-zero positive constant for any non-zero $a$.

\begin{table}[htb]
\centering
\caption{Estimated scaling exponent $\alpha$ of a variety of disordered hyperuniform systems, which are masked by both the Bernoulli and Debye masks. We note that in both cases, the masked systems are not hyperuniform anymore.}
\label{tab:scalingsII}
\begin{tabular}{c|c|c|c}
\toprule
 & gro. tru. & Bernoulli & Debye  \\ \hline
Class III & $0.5$ & 0.51$\pm$0.04 & -0.33$\pm$0.08   \\\hline
Class III & $0.7$ & 0.68$\pm$0.03 & -0.26$\pm$0.11  \\\hline
Class II & $1.0$ & 1.03$\pm$0.04 & -0.22$\pm$0.09   \\\hline
Class I & $1.3$ & 1.26$\pm$0.07 & -0.23$\pm$0.12   \\\hline
Class I & $1.5$ & 1.38$\pm$0.08 & -0.16$\pm$0.09   \\ \hline
Class I & $3.0$ & 2.79$\pm$0.13 & -0.18$\pm$0.12  \\
\hline
\hline
\end{tabular}
\end{table}


To verify these results, we numerically computed the structure factor of the same set of disordered hyperuniform point configurations with varying hyperuniformity exponent used for finite-window calculations, which are masked by both the Bernoulli and Debye masks. The numerically estimated exponents are given in Table II, which are compared to ground-truth values. It can be seen that for Bernoulli masks, one recovers a simple attenuation plus white-noise addition, which do not significantly change the estimated $\alpha$ values (within numerical tolerance). For the Debye mask, its \(\chi_M\) has non-negligible weight near \(k\approx0\) (e.g., large correlated mask features). Therefore, the small-\(k\) behavior of $S_{\mathrm{obs}}(\mathbf k)$ is strongly altered and the measured exponent is completely different from the original systems. 



Formally one could attempt to recover \(S(\mathbf k)\) by dividing the Fourier-domain relation by \(\chi_M\) (i.e. deconvolution). In practice this is ill-conditioned where \(\chi_M(\mathbf k)\) is small or zero (noise amplification). Regularized deconvolution or spectral-windowing methods are typically required. If the mask correlates with the point positions (selection bias), Eq.~\eqref{eq:raw-conv} does not factor and cross-terms appear. Those must be included explicitly and can produce additional nontrivial contributions (bias) to \(S_{\mathrm{obs}}\).

\section{Conclusions and Discussion}

In this work, we have developed a unified theoretical framework to quantify how finite observation windows and spatially correlated masks modify the measured structure factor of hyperuniform systems. By explicitly deriving the convolution relation between the observed structure factor $S_{\mathrm{obs}}(\mathbf{k})$ and the intrinsic structure factor $S(\mathbf{k})$, we have shown that any observation procedure that multiplies the density field in real space, whether through a compact window or a binary mask, inevitably redistributes spectral weight in Fourier space. This redistribution generates a leakage contribution that alters the small-wavenumber behavior of the measured spectrum. For generic hyperuniform systems with $S(k)\sim k^{\alpha}$, we demonstrated that finite windows impose a universal quadratic curvature in the lowest accessible wavenumbers, leading to an apparent $k^2$ scaling that is independent of the true exponent $\alpha$. Consequently, the intrinsic hyperuniform scaling can only be reliably extracted within the intermediate regime $1/L \ll k \ll q_c$, where $L$ is the linear size of the observation window and $q_c$ denotes the crossover scale beyond which the small-$k$ power law ceases to hold.

Our analysis further clarifies that the severity of measurement-induced distortion depends sensitively on the true exponent $\alpha$, the window size $L$, and the noise floor associated with finite sampling. In particular, for larger exponents ($\alpha \gtrsim 2$), the intrinsic signal $S(k)\sim k^\alpha$ decays rapidly at small $k$, making it increasingly susceptible to contamination by leakage and additive noise. This explains why exponent estimation for strongly hyperuniform systems can exhibit systematic downward bias unless sufficiently large observation windows or ensemble averaging are employed. In stealthy hyperuniform systems, where $S(k)$ possesses an exact spectral gap, the situation is even more stringent: all observed small-$k$ power originates from the convolution mechanism, and any apparent scaling behavior at low wavenumbers reflects the geometry and spectral density of the observation function rather than intrinsic density fluctuations.

The convolution formalism derived here also generalizes naturally to spatially correlated masks characterized by a spectral density $\chi_M(\mathbf{k})$. In this broader setting, the observed structure factor is governed by the interplay between $S(\mathbf{k})$ and $\chi_M(\mathbf{k})$, and hyperuniform signatures may be either suppressed or artificially induced depending on the low-$k$ behavior of the mask spectrum. This perspective highlights that incomplete or heterogeneous sampling can fundamentally alter spectral diagnostics of order. Therefore, reliable detection of hyperuniformity requires either careful control of the observation function or explicit correction via deconvolution or model-based fitting procedures.

We note that the present convolution framework also clarifies the feasibility of the inverse problem, i.e., estimating the intrinsic exponent $\alpha$ from finite-window measurements. A natural strategy is to compute an effective exponent $\alpha_{\mathrm{eff}}(L)$ for increasing window sizes and test for convergence. This approach can be successful when increasing $L$ opens a clear intermediate regime $1/L \ll k \ll q_c$, where the observed spectrum retains the intrinsic scaling $S(k)\sim k^\alpha$. However, this convergence is not guaranteed. If the intrinsic scaling range is narrow, if the exponent is large so that $S(k)$ is extremely small at accessible low $k$, or if leakage and finite-sample noise introduce a significant additive floor, $\alpha_{\mathrm{eff}}(L)$ may remain biased even for large windows. In stealthy systems, the small-$k$ signal from finite windows is entirely observation-induced and should not be interpreted as an intrinsic power law. For correlated masks, inverse recovery further depends on the low-$k$ behavior of the mask spectral density $\chi_M(k)$, which can preserve, suppress, or distort the intrinsic hyperuniform signature. Therefore, systematic variation of $L$ is a valuable diagnostic, but reliable extraction of $\alpha$ requires confirming the existence of a window-independent intermediate scaling regime.

Although the scaling arguments above were presented for a parent spectrum with a well-defined power-law regime $S(k)\sim k^\alpha$, the convolution framework does not require this assumption. It could be applied to systems with a scale-dependent local exponent
\begin{equation}
    \alpha_{\mathrm{loc}}(k)=\frac{d\ln S(k)}{d\ln k}.
\end{equation}
For such systems, the observed spectrum is a window-averaged version of the intrinsic spectrum over a spectral bandwidth of order $1/L$. The measured local exponent
\begin{equation}
    \alpha_{\mathrm{obs}}(k;L)=\frac{d\ln S_{\mathrm{obs}}(k;L)}{d\ln k}
\end{equation}
therefore represents an effective exponent determined by both the intrinsic scale dependence of $S(k)$ and the spectral width of the observation function. Increasing $L$ narrows the convolution kernel and can make $\alpha_{\mathrm{obs}}(k;L)$ approach $\alpha_{\mathrm{loc}}(k)$ at accessible $k$, but it does not by itself guarantee recovery of the strict asymptotic exponent $\alpha=\lim_{k\to0}\alpha_{\mathrm{loc}}(k)$. If $\alpha_{\mathrm{loc}}(k)$ evolves slowly and no clear crossover scale exists, the inverse problem is intrinsically ill-conditioned: finite-window measurements can characterize scale-dependent effective behavior, while extrapolation to the true asymptotic exponent requires additional assumptions about the functional form of $S(k)$.

For quasiperiodic structures, the convolution framework remains valid but must be interpreted differently. The intrinsic structure factor is a pure-point spectrum,
\begin{equation}
    S(\mathbf k)=\sum_{\mathbf G} I_{\mathbf G}\delta(\mathbf k-\mathbf G),
\end{equation}
so finite-window observation gives
\begin{equation}
    S_{\mathrm{obs}}(\mathbf k)
=
\sum_{\mathbf G}
I_{\mathbf G}
|\widetilde W(\mathbf k-\mathbf G)|^2 .
\end{equation}
Thus, each Bragg peak is broadened by the window power spectrum, and the measured signal is a superposition of broadened peaks rather than a smoothed continuous curve. When the reciprocal set is dense, this superposition can obscure any underlying scaling envelope, especially if the window broadening scale $\Delta k\sim 1/L$ is comparable to the spacing between relevant peaks. In such cases, the appropriate quantity for extracting scaling is not the pointwise value of $S_{\mathrm{obs}}(k)$, but rather peak-resolved or shell-integrated spectral weight \cite{Og17}, such as
\begin{equation}
    Z(K)=\sum_{0<|\mathbf G|<K} I_{\mathbf G}.
\end{equation}
The exponent associated with an intensity envelope can be recovered only when the window is large enough to resolve or statistically average over the relevant Bragg peaks. Moreover, this envelope exponent need not be identical to the strict hyperuniformity exponent, which is governed by the accumulation of total spectral weight near the origin. Therefore, for quasicrystals, the present formalism provides a useful forward model and a basis for finite-window correction, but inverse extraction of an exponent requires additional care.

Hyperuniformity in real systems is often identified in a coarse-grained or partial density field rather than in the total microscopic density \cite{chen2026machine, chen2024emergence}. The present framework applies to any such chosen density field, with the corresponding observation function determining the measured structure factor. Since the structural mechanisms that enhance hyperuniformity often involve local organization, clustering, or intra- and intermolecular correlations, finite windows may preserve the relevant low-$k$ suppression over an accessible intermediate range. Nevertheless, finite-window measurements cannot by themselves establish the strict zero-$k$ limit; they demonstrate scale-limited hyperuniform behavior whose interpretation depends on the chosen coarse-graining and observation scale.

Overall, our results provide quantitative criteria for identifying intrinsic hyperuniform scaling in realistic finite systems and offer practical guidance for interpreting numerical simulations, scattering experiments, and imaging-based measurements. By distinguishing genuine structural signatures from observation-induced artifacts, this framework contributes to a more robust and reproducible characterization of hyperuniform and stealthy states across physical, biological, and engineered systems. 
For a full simulation cell with periodic boundary conditions (PBCs), computing $S(k)$ only at box-commensurate wavevectors does not introduce the finite-window leakage discussed here; PBCs remove boundary discontinuities, although the accessible wavevectors remain discrete with $k_{\min}=2\pi/L_{\mathrm{box}}$. However, if finite subwindows, non-periodic subdomains, spatial masks, or non-commensurate wavevectors are used, the density is effectively multiplied by an observation function in real space, and the same convolution-induced leakage applies. Thus, PBCs do not remove leakage caused by subsequent windowing or masking; they only define the parent periodic spectrum.

\begin{acknowledgments}
This work was supported by the Army Research Office under Cooperative Agreement Number W911NF-22-2-0103. The author thanks Wenlong Shi for help with numerical calculations.
\end{acknowledgments}







\end{document}